# Crossover of magnetoresistance in the zerogap half-metallic Heusler alloy $Fe_2CoSi$

Y. Du,[1] G. Z. Xu,[1] X. M. Zhang,[1] Z. Y. Liu,[2] S. Y. Yu,[3] E. K. Liu,[1] W. H. Wang,[1, a)] and G. H. Wu[1]

[1]*Beijing National Laboratory for Condensed Matter Physics, Institute of Physics, Chinese Academy of Sciences, Beijing 100080, P. R. China*

[2]*State Key Laboratory of Metastable Material Sciences and Technology, Yanshan University Technology, Qinhuangdao 066004, P. R. China*

[3]*School of Physics, Shandong University, Jinan 250100, P. R. China*

This work reports on the band structure and magneto-transport investigations of the inverse Heusler compound $Fe_2CoSi$. The first-principles calculations reveal that $Fe_2CoSi$ has a very peculiar band structure with a conducting property in the majority spin channel and a nearly zero bandgap in the minority spin channel. The synthesized $Fe_2CoSi$ sample shows a high-ordered inverse Heusler structure with a magnetic moment of 4.88 $\mu_B$ at 5 K and a high Curie temperature of 1038 K. With increasing temperature, a crossover from positive to negative magnetoresistance (MR) is observed. Complemented with the Hall effect measurements, we suggest the intriguing crossover of MR can be ascribed to the dominant spin carriers changing from the gapless minority spin channel to the majority spin channel at Fermi level.

The magnetoresistance (MR) effect in ferromagnets has been one of the important subjects in experimental and theoretical studies over past years [1-4]. Due to spin-dependent scattering, ferromangnets usually exhibit a negative MR and saturates at high fields[5-7]. A positive and non-saturating linear MR in ferromagnets is therefore interesting for both the fundamental magnetotransport phenomena and magnetic sensor applications. So far, the positive and non-saturating MR has been reported in many systems, such as the rare earth metal multilayers Dy/Sc[8] and slightly doped silver chalcogenides, $Ag_{2+\sigma}Se$ and $Ag_{2+\sigma}Te$[9,10]. They have been interpreted as the consequence of either (i) the suppression of interfacial reflectivity cased by magnetization of the interfacial alloy; or (ii) the quantum MR proposed by Abrikosov[11], where only the lowest Landau level remains occupied[12]. Recently, a crossover of MR from negative to positive was observed in half-metallic Heusler compound NiMnSb[13] and spin gapless inverse Heusler compound $Mn_2CoAl$[14], where the resistance displayed a positive dependence on the magnetic field at low temperatures, without any signs of saturation. These results indicate the electric structure around the Fermi level ($E_F$) plays a key ingredient in understanding the remarkable MR variations in ferromagnets. In this letter, we report on the crossover of MR from negative to positive with increasing temperature in the inverse Heusler compound $Fe_2CoSi$. Based on first principles calculations (see Figure 1), we found that the $Fe_2CoSi$ shows a zerogap half-metallic valence band spectrum close to the Fermi energy. Similar to the spin gapless semiconductor, the crossover behavior of MR can be attributed to the unique band structure of this kind of material.

Figure 1 shows the spin resolved band structure and density of states (DOS) for $Fe_2CoSi$. The calculations were performed using the pseudopotential method with plane-wave-basis set. The electronic exchange correlation energy was treated under the generalized gradient approximation for the experimental lattice constant of $Fe_2CoSi$, 0.564 nm (see below). We used 120 points in the irreducible wedge of the Brillouin zone of the primitive cell and a plane-wave cut-off energy of 500 eV. These parameters ensure good convergences for total energy. In Figure 1, the spin character of the bands is indicated by different colors. In the majority states, several band cross the $E_F$ in the G-X and G-L directions. These bands are strongly dispersing and contribute mainly to the large density in the majority states. The situation is different in the minority channel. Few flat bands appear to be just closed as a result of bands touching the $E_F$ at high symmetry (G, X) points. This particular band structure suggests that there is a strong difference in the conductivity for the minority and the majority electrons. Following the classification of Wang et al.[15], the zero bandgap behavior of the minority electrons converts the gapless semiconductor[16] into a gapless half-metallic ferromagnet. Nevertheless, the calculated spin polarization at the $E_F$ is still 100%, and the magnetic moment is calculated to be 5 $\mu_B$, as observed experimentally (see inset of Fig. 2(b)). In the following, we will show a crossover of magnetoresistence from positive to negative behavior in the proposed gapless half-metallic Heusler $Fe_2CoSi$ alloy.

The nominal composition of $Fe_2CoSi$ samples were prepared by arc melting pure elements under argon atmosphere in a water-cooled Cu crucible. The ingots were re-melted several times and subsequently melt-spun using a single-roll melt-spinner at a wheel angular speed of 60 rad/s. The structural examinations were carried out at room temperature with powder X-ray diffraction (XRD) techniques using Cu-Ka radiation. Figure 2 (a) shows the XRD pattern of $Fe_2CoSi$. Indexing the characteristic reflections, we find that the fabricated sample is face-center-cubic (fcc) phase with $Hg_2CuTi$ structure (see right inset of Fig. 2(a)). The lattice parameters found from the XRD patent are $a$=0.564 nm. The left inset of Fig. 2 (a) shows the XRD superlattice reflections measured by step-scan. The appearance of superlattice (111) and (200) indicates that our samples have a highly ordered structure. Figure 2 (b) shows the temperature dependence of the zero-field resistivity. As expected, the resistivity decreases with decreasing temperature in the temperature range. Here we should point out that in many Heusler alloys[17,18], especially including Mn atoms, the resistivity always shows a local minimum at low temperatures with the resistivity increasing slightly toward lower temperature. However, this type of low-temperature anomaly is not observed in our high-ordered $Fe_2CoSi$. Moreover, below 100 K, the resistivity decrease with temperature is well fitted by a power law $T^\alpha$ with $\alpha = 3.1$, close to the value 3, which is experimental confirmed in half-metallic ferromagnets[19,20]. The temperature dependence of the magnetization is shown in the Fig. 2 (c). The measurements were performed at a field of 500Oe. The cause of the dramatic decrease of the magnetization at 1038 K as marked by the arrow is the Curie temperature of $Fe_2CoSi$, which is the highest value for inverse Heusler compounds being measured up to now. A superconducting quantum interference device (SQUID, Quantum Design

Inc.) was used to determine the saturation magnetic moment in the low temperature of 5K. The results of the magnetization measurements are displayed in the inset of Figure 2 (c). The magnetization at an induction field of 5 T corresponds to a magnetic moment of about 4.88 $\mu_B$ in the unit cell, which is well consistent with the calculated value of 5 $\mu_B$.

From the application point of view, the MR is an important fact for particular interest. The MR was measured by the standard four-terminal method in Physical Property Measurement System (PPMS, Quantum Design Inc.), applying a direct current (DC) of 5000 μA in the direction that parallel to the magnetic field. In Figure 3, we show the results of normalized MR as a function of magnetic field $H$ at different temperatures. The MR was calculated from the resistivity ($\rho_{xx}$) with and without a magnetic field and defined as, $MR = [\rho_{xx}(0) - \rho_{xx}(H)]/\rho_{xx}(0)$. The values of MR at temperatures below 300 K are positive and increase almost linearly with the magnetic field. That is, the relation between MR and magnetic field can be described by $MR \propto \beta(T)H$, where $\beta(T)$ is a coefficient and could be temperature dependent. The negative value of the *MR* indicates that the spin dependent scattering is dominating at temperatures above 200 K. Such kind of temperature dependence, with changed sign of the MR at a given temperature, could arise only from different temperature dependence of the positive and negative components to the total MR signal. At temperatures below 200 K, the negative signal dominates over the positive signal, whereas above this temperature the positive signal dominates over the negative one.

The question arises as to the origin of the positive MR at low temperature in our Fe$_2$CoSi sample. As we have mentioned above, due to the spin dependent scattering, the MR in ferromagnets usually display a negative sign and its amplitude should increase with increasing temperature. The appearance of a positive MR is usually considered as the contribution of Lorenz force to resistivity in the presence of magnetic field[21]. However, we can rule out such a scenario because the magnitude of the positive MR induced by Lorentz force is found to be proportional to square of the magnetic field; however we found the MR of Fe$_2$CoSi is nearly linear and not proportional to $H^2$. Very recently, spin gapless semiconductor (SGS) in Heusler has been predicted theoretically by Wang et al.[15], and experimentally Ouardi et al.[14] has confirmed that the sign of MR in SGS Mn$_2$CoAl Heusler alloy is positive at low temperatures and changes to negative with increasing temperature. It is worth noting that, as shown in Fig. 1, the Fe$_2$CoSi presents a gapless semiconductor-like band as marked by the circle at $E_F$ in spin-down band, and the other band passes though the $E_F$.

Figure 4 shows Hall resistivity $\rho_{xy}$ as a function of magnetic field $H$ at various temperatures. The anomalous Hall conductivity $\sigma_{xy} = \dfrac{\rho_{xy}}{\rho_{xx}^2}$ was obtained from the magnetic-field-dependent transport measurements. We found that, the anomalous Hall conductivity $\sigma_{xy}$ has a low value of 67.06 S/cm at 10 K, which has the same magnitude than the SGS Mn$_2$CoAl[14]. The temperature-dependent MR percentage rates and the Ordinary Hall Effect (OHE) coefficient R$_O$ was plotted in Fig. 4 (b). We found that at temperatures below 200 K, the sign of OHE coefficient R$_O$ is positive, while the sign changed to negative at temperatures higher than 200 K. Above the saturation magnetic field of 2 T, it is found that a change of sign in $R_O$ emerge at around 200 K and the approximate extent of temperature is consistent with the MR changed the sign. Moreover, as shown in Fig. 4 (c), the anomalous Hall resistivity $\rho_{xy}$ as a function of longitudinal resistivity $\rho_{xx}$ can divide into two regions, this function does not possess continuity and the slopes of these two sections are completely different. Though our analysis, the present results demonstrates that there may be the competition between two kinds of carrier conductive mechanism, it will be discussed below. Above 150K, the experiments to measure the relationship between $\rho_{xy}$ and $\rho_{xx}$ generally assumed to be of the linear law form, i.e., $\rho_{xy} \sim \rho_{xx}$, the main source of the AHE currents was skew scattering from impurities caused by the spin-orbit interaction (SOI)[22,23].

On the analogy of the above-mentioned spin gapless semiconductor, in Figure 4, we show the brief density of states (DOS) schemes for three different situations. Unlike ordinary half-metal and spin gapless semiconductor, for $Fe_2CoSi$, the spin-down channel is gapless, while the spin-up channel is conducting. Due to the spin dependent scattering, the spin-up conduction electrons in $Fe_2CoSi$ Heusler alloy result in MR decrease with increasing magnetic field. However, no threshold energy is required to move electrons from occupied states to empty states in this zero bandgap and unique transport properties that the sign of MR is positive emerge in the spin down gapless semiconductor-like band, influenced by magnetic fields. It is possible that the competition between these two mechanisms lead to the appearance of positive magnetoresistence properties. With increasing temperature, the effect of spin dependent scattering will gradually increase. At the same time, zero bandgap will be destroyed by thermal excitation and the special band structure will disappear, so that finally the sign of MR changes to negative at room temperature.

In summary, ab initio calculations suggested that the Heusler compound $Fe_2CoSi$ exhibit conducting properties in one of the spin channels and a zero bandgap in the other. This sample is pure fcc phase with $Hg_2CuTi$ structure, and saturation magnetic moment measured by SQUID is 4.88 $\mu_B$ at 5K compared with theoretical calculating value of 5.0 $\mu_B$. The curie temperature of 1038K makes it suitable for applications. It has been shown by experiments that temperature-dependent resistivity of $Fe_2CoSi$ is well fitted by a power law $T^\alpha$ with $\alpha = 3.1$. Interestingly, a crossover of magnetoresistance from negative to positive was observed in $Fe_2CoSi$. In high fields, it is positive and non-saturating at low temperatures, but the sign becomes negative at high temperatures. Moreover, the change signs of Ordinary Hall Effect (OHE) coefficient and piecewise function type $\rho_{xy} \propto \rho_{xx}$ curve imply that there may be the competition between two kinds of carrier conductive mechanism. This kind of material is defined as zero-gap half metal and it may possess unique MR behavior similar to spin gapless semiconductor.

This work is supported by the National Basic Research Program of China (973 Program 2012CB619405) and National Natural Science Foundation of China (Grant Nos. 51071172, 51171207 and 51171206)

**Figure captions:**

FIG. 1. (Color online) Spin-resolved electronic structure of inverse Heusler $Fe_2CoSi$. Shown are the band structures for the minority (a) and majority (c) electrons together with the density of states (b). The $Fe_2CoSi$ presents a gapless semiconductor-like band as marked by black circle at $E_F$.

FIG. 2. (Color online) (a) Powder X-ray diffraction of $Fe_2CoSi$ sample. The inset shows the crystal structure. (b) Temperature dependence of resistivity of $Fe_2CoSi$ measured at a zero field. The inset shows the resistivity at the low-temperature region and the data are fitted to the $T^{3.1}$ law. (c) Temperature dependence of magnetization measured in a field of 1 T. The inset shows the magnetization curve measured at 5 K.

FIG. 3. (Color online) The magnetoresistance (MR), MR=$[\rho(H)-\rho(0)]/\rho(0)$, as a function of magnetic field measured at different temperatures. All data were taken in the magnetic field parallel to the sample plane. The positive MR can be fitted qualitatively by $\beta(T)H$, where the coefficient $\beta(T)$ could be temperature dependent.

FIG. 4. (Color online) (a) Field-dependent Hall resistivity $\rho_{xy}$, as a function of magnetic field $H$ at various temperatures. (b) The temperature-dependent MR percentage rates (black solid point) and the ordinary Hall Effect (OHE) coefficient $R_O$ (blue solid point). (c) The anomalous Hall resistivity $\rho_{xy}$ as a function of longitudinal resistivity $\rho_{xx}$.

FIG. 5. (Color online) The schematic density of states n(E) as a function of energy E is shown for (a) a half-metallic ferromagnet, (b) a spin gapless semiconductor, and (c) a zerogap half-metallic ferromagnet. The occupied states are indicated by filled areas. Arrows indicate the majority ($\uparrow$) and minority ($\downarrow$) states.

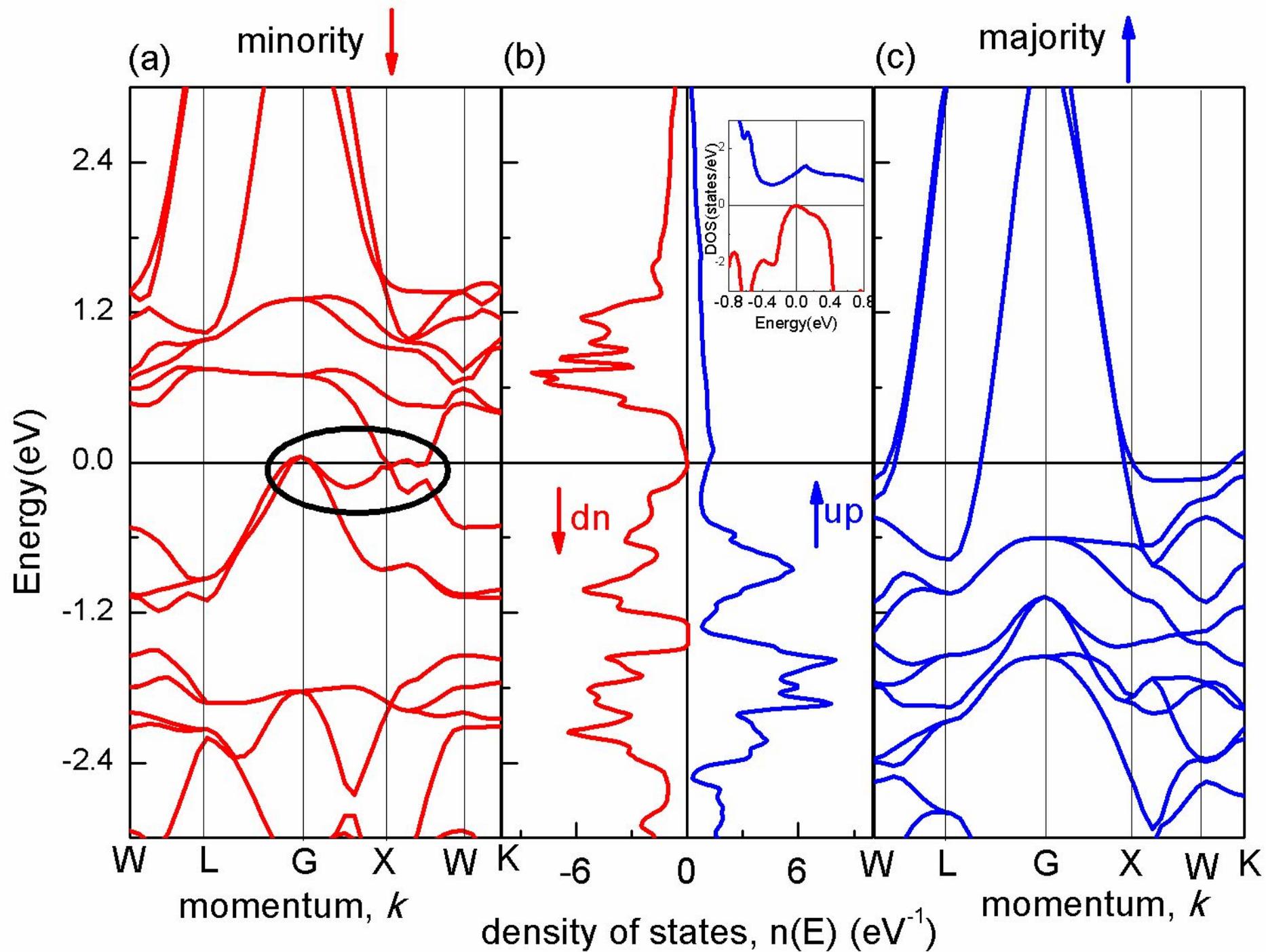

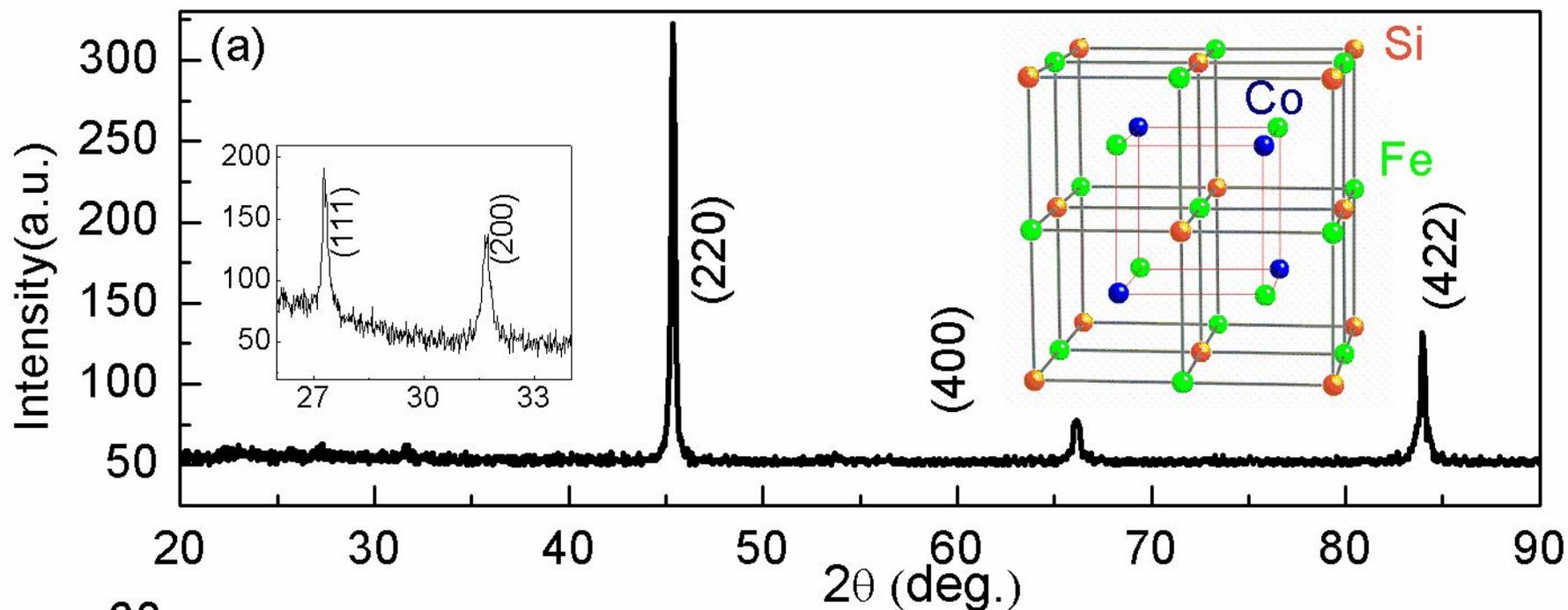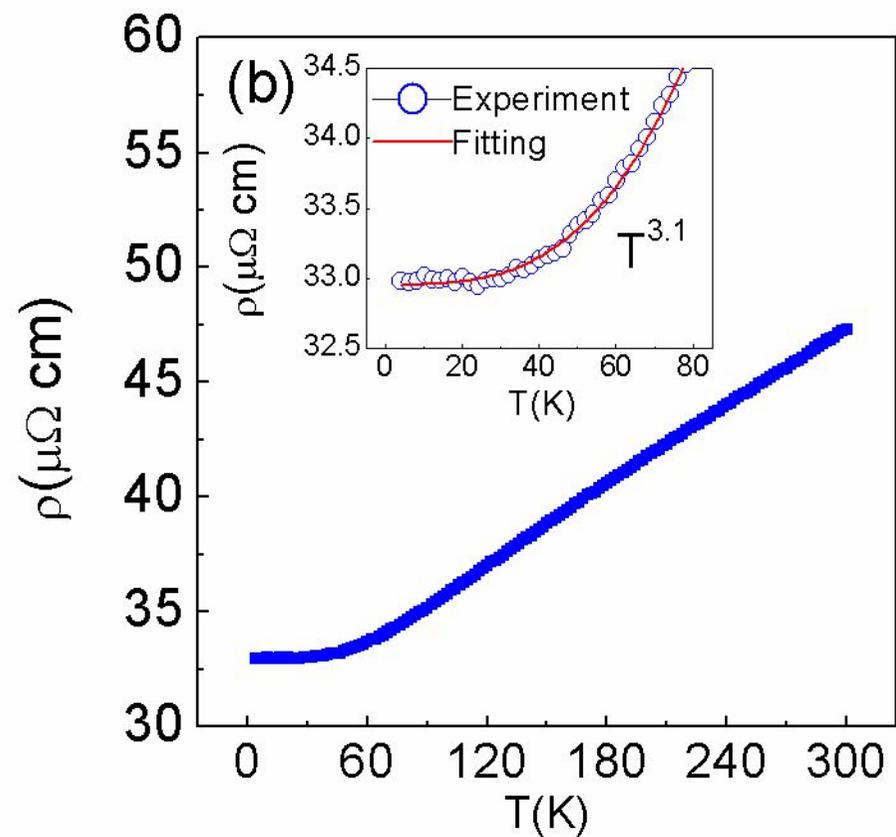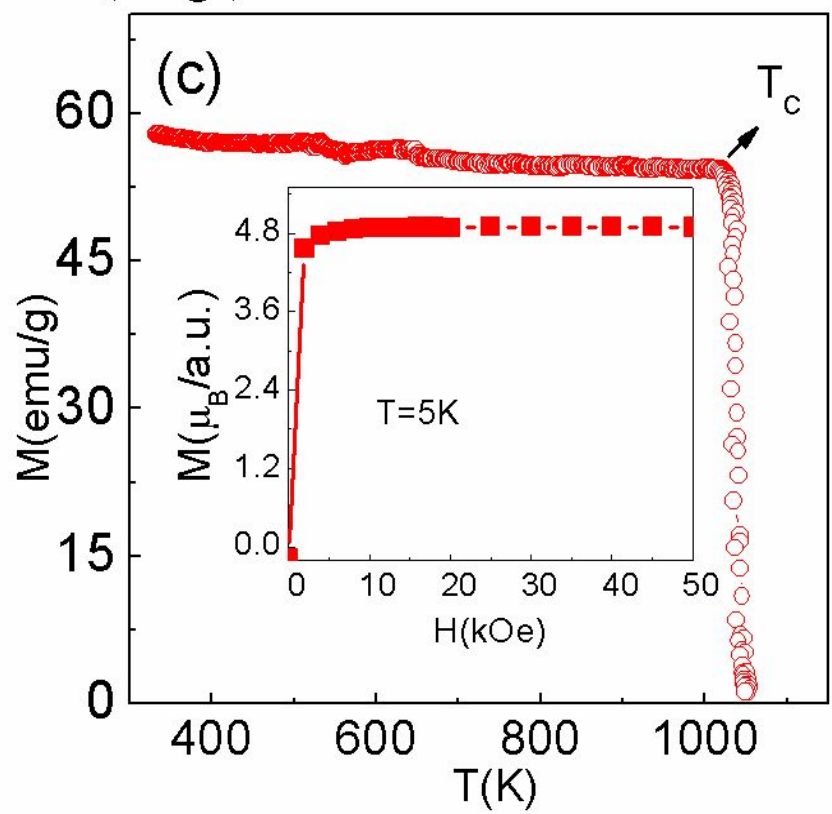

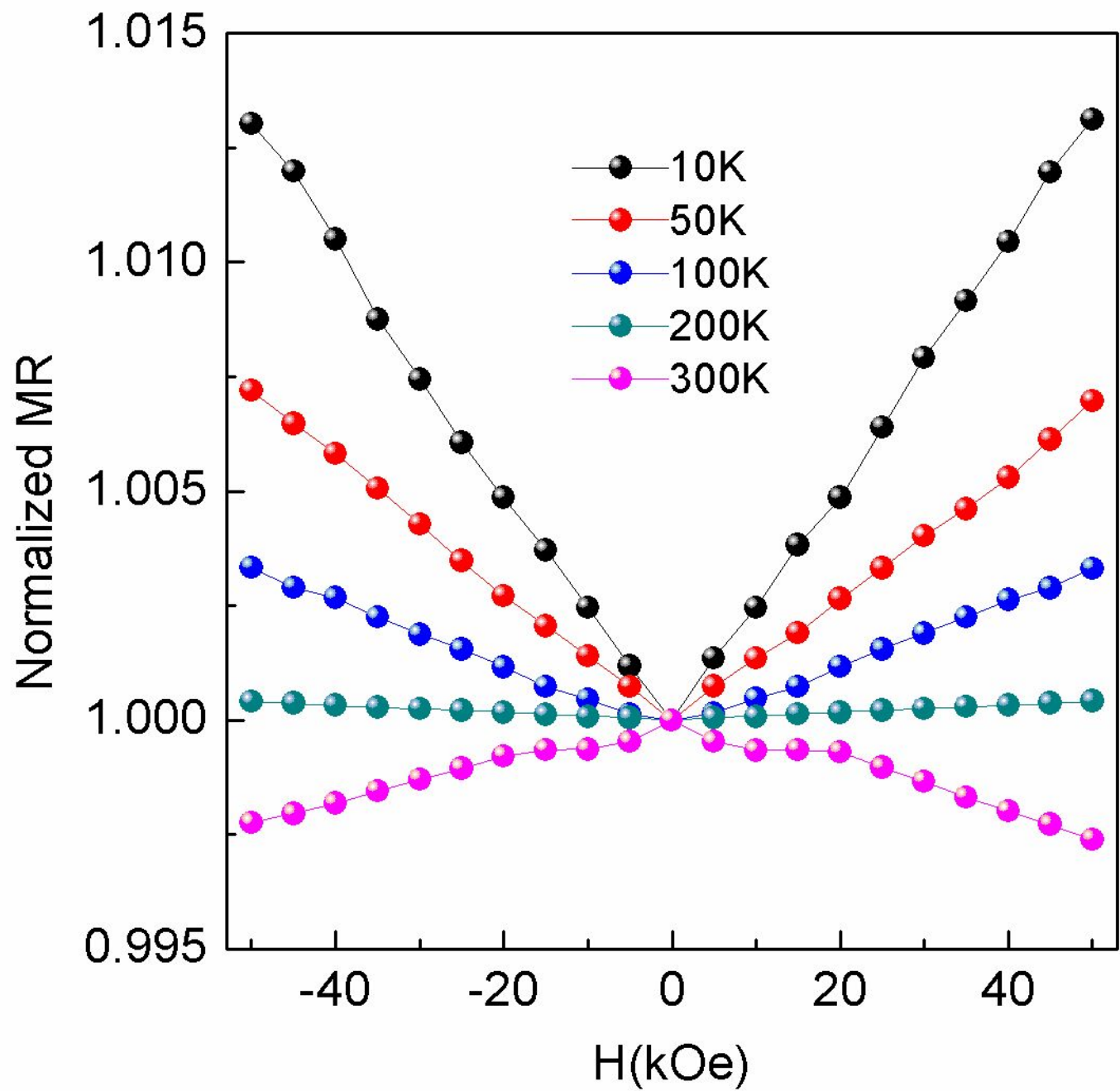

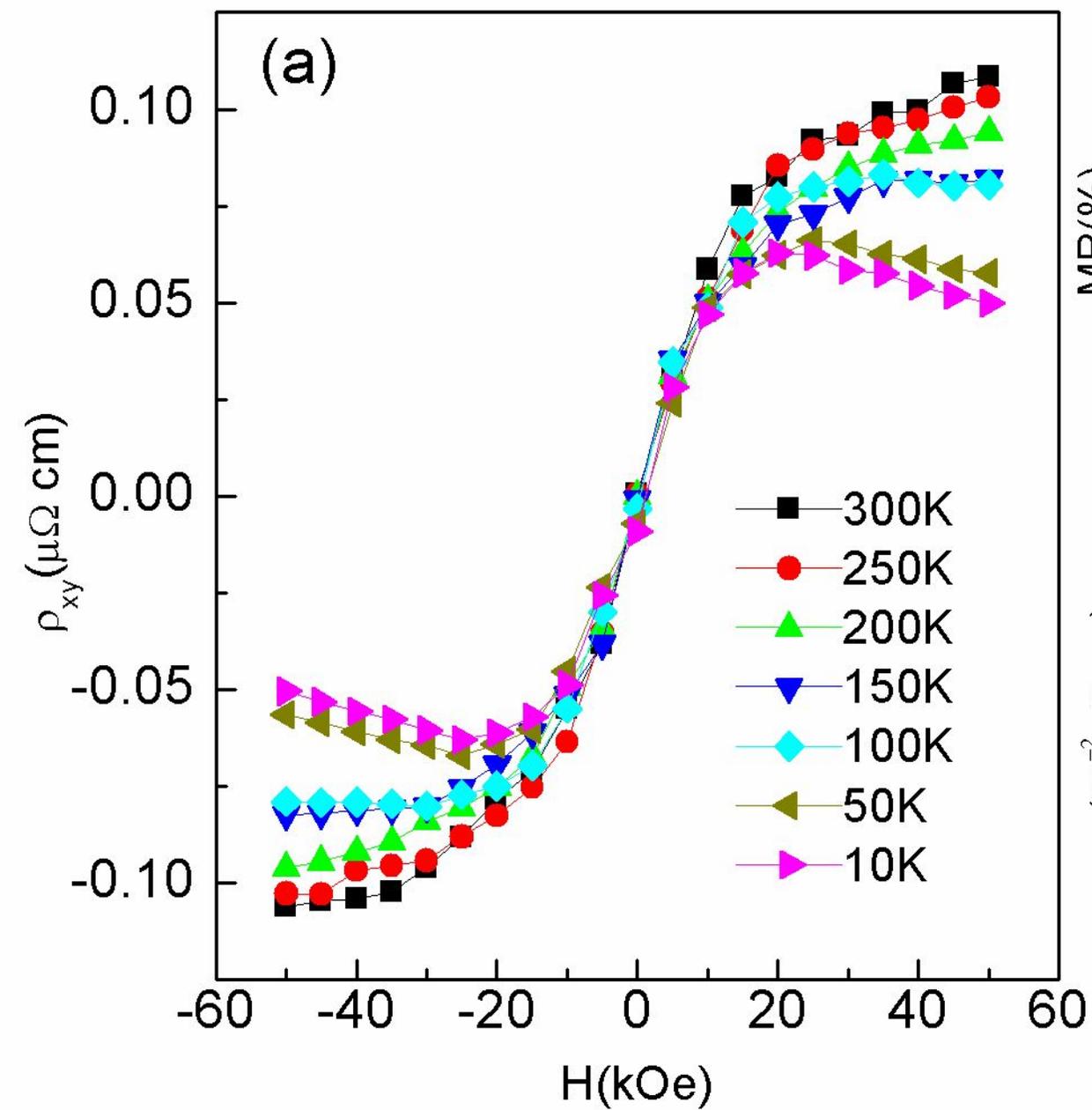

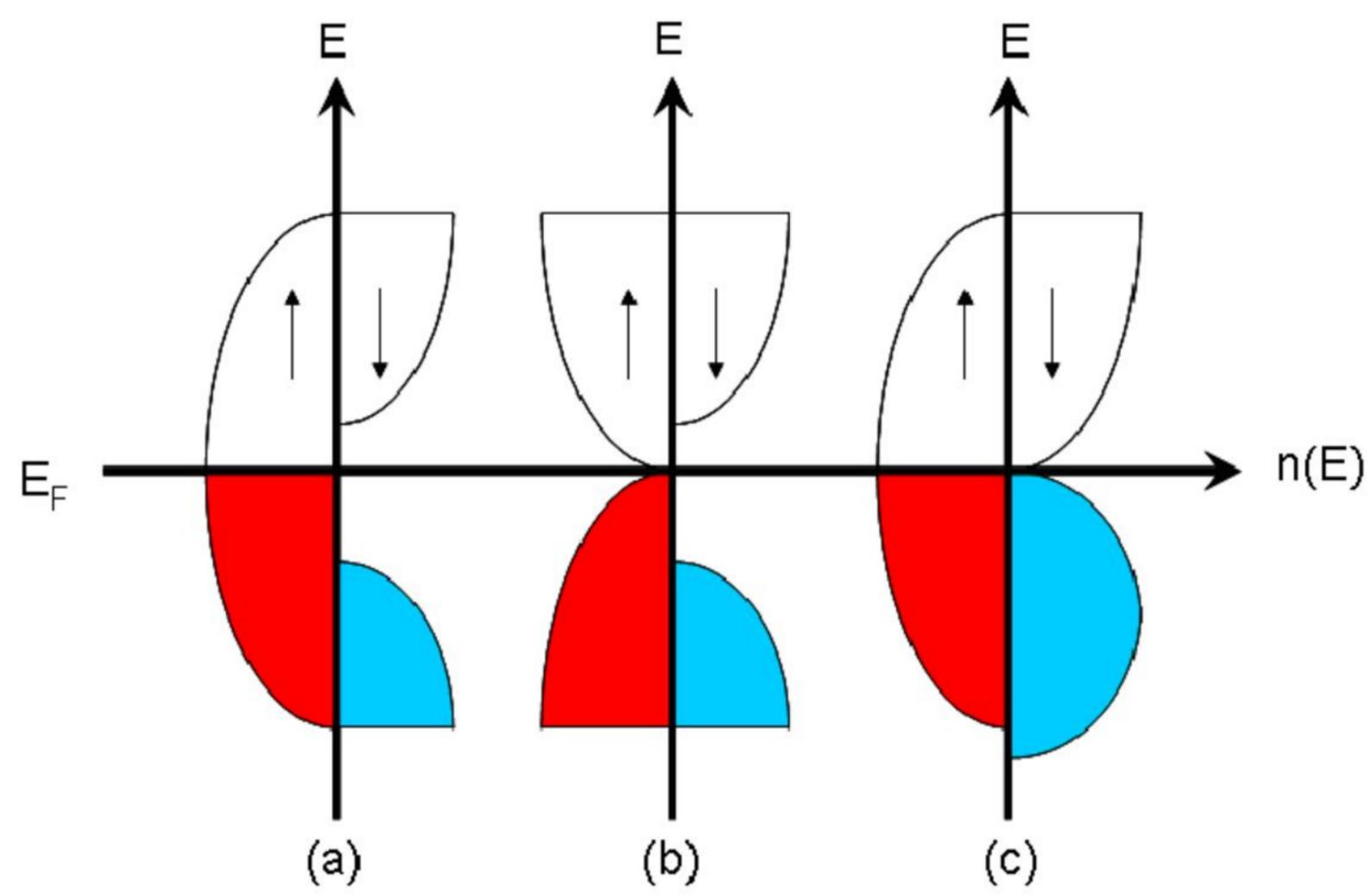